\documentclass[manuscript]{aastex}
\newcommand{\lum}{erg s$^{-1}$}
\newcommand{\flux}{erg cm$^{-2}$ s$^{-1}$}

\newcommand{\ltsima}{$\; \buildrel < \over \sim \;$}
\newcommand{\simlt}{\lower.5ex\hbox{\ltsima}} 
\newcommand{\gtsima}{$\; \buildrel > \over \sim \;$}
\newcommand{\simgt}{\lower.5ex\hbox{\gtsima}} 

\newcommand{\swift}{{\it Swift}}
\newcommand{\spitzer}{{\it Spitzer}}
\newcommand{\chandra}{{\it Chandra}}
\newcommand{\xmm}{{\it XMM-Newton}}

\newcommand{\euve}{{\it EUVE}}
\newcommand{\galex}{{\it GALEX}}
\newcommand{\rosat}{{\it ROSAT}}
\newcommand{\sax}{{\it BeppoSAX}}
\newcommand{\asca}{{\it ASCA}}
\newcommand{\suzaku}{{\it Suzaku}}
\newcommand{\hst}{{\it HST}}
\newcommand{\fg}{{\it g}}
\newcommand{\fr}{{\it r}}
\newcommand{\fii}{{\it i}}
\newcommand{\fz}{{\it Z}}
\newcommand{\fJ}{{\it J}}
\newcommand{\fH}{{\it H}}
\newcommand{\fU}{{\it U}}

\newcommand{\fV}{{\it V}}
\newcommand{\fB}{{\it B}}
\newcommand{\fW}{{\it UVW1}}

\newcommand{\heasarc}{{\it HEASARC}}

\shorttitle{TDE in a Host Galaxy with an IMBH}
\shortauthors{Donato et al.}

\begin{document}

\title{A Tidal Disruption Event in a Nearby Galaxy \\
       Hosting an Intermediate Mass Black Hole}

\author{D. Donato\altaffilmark{1,2} and S. B. Cenko\altaffilmark{3}}
\email{davide.donato-1@nasa.gov}

\author{S. Covino\altaffilmark{4}, E. Troja\altaffilmark{1}, T. Pursimo\altaffilmark{5}, C. C. Cheung\altaffilmark{6}, O. Fox\altaffilmark{3}, A. Kutyrev\altaffilmark{7}, S. Campana\altaffilmark{4}, D. Fugazza\altaffilmark{4}, H. Landt\altaffilmark{8}, N. R. Butler\altaffilmark{9}}


\altaffiltext{1}{CRESST and Astroparticle Physics Laboratory NASA/GSFC, Greenbelt, MD 20771, USA}
\altaffiltext{2}{Department of Astronomy, University of Maryland, College Park, MD 20742, USA}

\altaffiltext{3}{Astrophysics Science Division, NASA/GSFC, Mail Code 661, Greenbelt, MD 20771, USA}
\altaffiltext{4}{INAF, Osservatorio Astronomico di Brera, via E. Bianchi 46, 23807 Merate (LC), Italy}
\altaffiltext{5}{Nordic Optical Telescope, Apartado 474, 38700 Santa Cruz de La Palma, Spain}
\altaffiltext{6}{Space Science Division, Naval Research Laboratory, Washington, DC 20375-5352, USA}
\altaffiltext{7}{Observational Cosmology Laboratory, NASA/GSFC, 8800 Greenbelt Road, Greenbelt, MD 20771-2400, USA}
\altaffiltext{8}{Department of Physics, Durham University, South Road, Durham DH1 3LE, UK}
\altaffiltext{9}{School of Earth and Space Exploration, Arizona State University, Tempe, AZ, USA}


\begin{abstract}
We report the serendipitous discovery of a bright point source flare in the 
Abell cluster 1795 with archival \euve\ and \chandra\ observations. 
Assuming the \euve\ emission is associated with the \chandra\ source, the X-ray
0.5--7 keV flux declined by a factor of $\sim 2300$ over a time span of 6 years,
following a power-law decay with index $\sim 2.44 \pm 0.40$.  The \chandra\
data alone vary by a factor of $\sim 20$. The spectrum is well fit by a 
blackbody with a constant temperature of $kT \sim 0.09$ keV ($\sim 10^6$ K). The
flare is spatially coincident with the nuclear region of a faint, inactive 
galaxy with a photometric redshift consistent at the one sigma level with the 
cluster ($z = 0.062476$). We argue that these properties are indicative of a 
tidal disruption of a star by a black hole with $log (M_{\rm BH}/M_{\odot}) \sim 
5.5 \pm 0.5$. If so, such a discovery indicates that tidal disruption flares may
be used to probe black holes in the intermediate mass range, which are very 
difficult to study by other means.
\end{abstract}


\section{Introduction} \label{intro}

There now exists compelling evidence that most, if not all, massive bulge 
galaxies harbor super-massive black holes (SMBH, $M_{\rm BH} \ge 10^6 M_{\odot}$) 
in their nuclei \citep[e.g.,][]{kor95}. During their lifetime those galaxies can
experience a phase of high activity when the gas is rapidly falling into the BH
through an accretion disk \citep[e.g.,][]{hop06}. But even in the 
subsequent quiescent phase, the BH can be fed by stars whose orbits are too 
close to escape
its large gravitational potential well \citep{fra76}. Such tidal disruption 
events (TDE) can lead to bright X-ray/UV flares as a fraction of the disrupted 
material accretes onto the BH \citep{ree88}. In addition to the X-ray/UV 
emission from the disk, radio emission has now been detected in several TDE 
candidates. This is interpreted as synchrotron emission from electrons 
accelerated by a relativistic jet formed during the disruption process 
\citep{blo11,zau11}. Indeed, only a few cases of a TDE accompanied by the birth
of the relativistic jet have been discovered: In GRB 110328A/Sw J1644+57 
\citep{bur11,lev11,zau11} and Sw J2058+05 \citep{cen12} the radio emission has 
been observed simultaneously to the X-ray discovery with the \swift\ satellite, 
while a radio source has been observed both in the nucleus of the galaxy IC 3599
and at the position of the transient RX J1420.4+5334 nine and 22 years after the
initial X-ray detection, respectively \citep[][and references therein]{bow13}.

Before these \swift\ discoveries, the first TDE candidates were identified in 
archival X-ray data, either in the R\"{o}ntgensatellit satellite 
all-sky survey \citep[\rosat,][]{gru95,bad96,kom99,gre00,don02,cap09}, in the 
X-ray Multi-Mirror \xmm\ slew survey \citep[\xmm,][]{esq08,esq10}, or in 
observed fields by \xmm\ and \chandra\ \citep{kom04,mak10,lin11,sax12}. 
More recently, other candidates have been found as a result of ongoing, 
real-time surveys. Some examples include transients discovered by the Galaxy 
Evolution Explorer \citep[\galex,][]{gez06,gez08,gez09}, in the Sloan Digital 
Sky Survey \citep[SDSS,][]{kom08,van11}, in the Pan-STARRS1 Medium Deep 
Survey \citep{gez12}, and by the Palomar Transient Factory 
\citep[PTF,][]{cen12b}.

In almost all these cases, the event occurred in a normal galaxy (i.e., 
without Seyfert activity); the flaring source reached an X-ray/UV luminosity of 
$L > 10^{42}$ erg s$^{-1}$ and faded by at least 1-2 orders of magnitude
on time scales of months/years; and, the spectral energy distribution (SED) was 
characterized by a black-body with a temperature $T$ \gtsima $10^5$ K, as 
expected from an accretion disk \citep{kom99,esq07,cap09,mak10}. It has been 
argued \citep{ree88,eva89,phi89} that the mass accretion rate should follow a $
\sim t^{-n}$ power-law where $n = 5/3$; however, more recent analytic works and 
hydrodynamical simulations \citep{str09,str11,lod09,lod11,gui13} suggest that 
the flux in any given band may deviate from this simple power-law. At early 
times the slope is expected to be flatter while at later times the slope 
asymptotes to $\sim2.2$ for approximately half of the stellar disruptions. 
\citet{gui13} consider the possibility that the star is only partially 
disrupted, with the stellar core surviving the encounter and with the stellar 
outer gas becoming bound to the BH. In all of these cases, the index is steeper 
than $n = 5/3$.

Observationally, the typical X-ray light curve is poorly sampled and the slope 
of decline is consequently not well constrained. In a few cases when the event 
is extensively monitored \citep[e.g., Sw J2058+05,][]{cen12} or is detected by 
chance on multiple occasions \citep{cap09}, a steep decline with $n \sim 2.2$ is
observed. The two \swift\ discoveries (Sw J1644+57 and Sw J2058+05) show also 
significant variation on relatively short time scales (minutes to days). 
On timescales of a few hours, the X-ray flux of these two transients 
changed by a factor of 100 and 1.5, respectively.

The hosts in nearly all the studied cases have been either a quiescent or a
star-forming galaxy. Morphologically they range from being ellipticals/S0 to 
spirals with, typically, an evident bulge. The common feature among the hosts is
the estimated BH mass: regardless of the luminosity scaling relation used to 
infer it, the BH is thought to be super-massive ($M_{\rm BH}/M_{\odot} \sim 
10^6-10^7 $). Only in a few cases an intermediate mass BH (IMBH, $10^2 \simlt 
M_{\rm BH}/M_{\odot} \simlt 10^6$) has been proposed by some authors. However, 
all of these cases differ from previous traditional TDEs because the disrupted 
object is not a main sequence star: a white dwarf for Swift J1644+57 
\citep{kro11}, the gamma--ray bursts GRB 060218 \citep{shc12} and GRB 060614 
\citep{lu08}, a flare in an extragalactic globular cluster \citep{irw10} 
and a super-Jupiter object for IGR J12580+0134 \citep{nik13}.
 
In this work we present the serendipitous discovery of an extremely bright point
source in archival observations with the Extreme Ultraviolet Explorer (\euve) 
and \chandra\ of the field of the moderately rich (richness class 2) cluster 
Abell 1795 \citep[A1795, $z = 0.062476$,][]{hil93}. The large X-ray flux 
variation (with a light curve characterized by a power-law decay), together with
the shape of the \chandra\ spectra suggests that this is a classical TDE, while 
the characteristics of the putative host galaxy identified in optical and 
infrared observations further suggest that the host is harboring an IMBH. 
In the final stages of preparation of this manuscript, an independent discovery
of this source was reported by \citet{mak13}. While these authors came to 
largely similar conclusions regarding the origin of the transient, in this
work we present 1) a more detailed light curve analysis based on very recent 
simulation studies to support our interpretation of the transient nature as a 
tidal event; 2) stringent limits on the jet emission using extensive archival 
analysis of VLA data; 3) new broadband photometric data and significantly deeper
spectroscopic limits that better constrain the characteristcs of the host
galaxy.

The paper is organized as follows: In Sect.~2 we present the data reduction and 
analysis of new and archival observations, in Sect.~3 we show the evidence 
supporting the TDE scenario and in Sect.~4 we discuss the results. A summary is
given in Sect.~5. Throughout the paper, a concordance cosmology with $H_{\rm 0} =
71$ km s$^{-1}$ Mpc$^{-1}$, $\Omega_{\Lambda}$ = 0.73, and $\Omega_{\rm m}$ = 0.27 
\citep{spe03} is adopted. Quoted errors are 90\% confidence levels for the 
X-ray analysis results and 68\% in all the other cases. All the upper limits
are at the 3$\sigma$ level unless stated otherwise.

\section{Observations and Data Analysis} \label{obs}

Since its launch in 1999, \chandra\ observed the field of the galaxy cluster 
A1795 several times. In the first pointing taken in 1999 December, a very bright
point-like source, which we refer to hereafter as CXO J1348, was detected 
0.7\arcmin\ west of the cD elliptical galaxy located at the center of the 
cluster. In the following four years, \chandra\ re-observed the same field five 
times (see Table~\ref{xray} for a detailed description of the \chandra\ 
pointings). In these observations the source is detected but with a declining
intensity. From 2005 to 2012, the transient position fell within the \chandra\ 
field of view 20 additional times, but no emission was detected at this 
position.

Intrigued by this behavior, we searched the archives of other satellites and 
telescopes and in the literature. We found that a few months before the first 
\chandra\ observation, a target of opportunity (ToO) was granted by the Advanced
Satellite for Cosmology and Astrophysics (\asca) 
team to observe a giant outburst from a galaxy in A1795 discovered in archival 
\euve\ data \citep{bow99,bon01}. This outburst appeared for the first time in 
an \euve\ observation performed in 1998 March 27 and disappeared in the 
following months. \citet{bow99} reported that the radial profile of the UV 
emission was consistent with being produced by a point source. The \asca\ ToO 
was triggered in coordination with further \euve\ follow-up observations to 
determine its presence in the UV/X-rays and its nature. Unfortunately, the large
\asca\ point spread function (PSF) and the bright intensity of the galaxy 
cluster in the X-ray prevented any detection, and no further action was taken. 
The point source was never detected again by \euve, suggesting a transient
origin.

We checked for additional data recorded by other X-ray satellites spanning
the 1992-2004 time range. The \rosat\ data provide useful timing constraints: 
four observations were obtained with the High Resolution Imager from 1992 July 
25 to 1997 July 23. For all of them we did not detect any X-ray emission at 
the position of CXO J1348. An upper limit in the 0.1--2.4 keV range for 
the count rate of the last observation (lasting 8.8 ksec) can be set at 
$< 4.42 \times 10^{-3}$ ct s$^{-1}$. Assuming a black-body spectrum with $kT = 
0.09$ keV (an average value obtained fitting the \chandra\ spectra of the 
observations taken in the 1999-2002 period; see Sect.~2.1.2) and Galactic 
absorption \citep[$N_{\rm H,Gal} = 1.19 \times 10^{20}$ cm$^{-2}$,][]{kab05}, 
we estimate with \verb+WebPIMMS+ an unabsorbed flux upper limit, 
extrapolated to the 0.5--7 keV range, of $\sim2.9 \times 10^{-14}$ \flux\ for 
the observation in 1997.

Data from other X-ray satellites, such as \xmm, \sax, \swift, and \suzaku, were 
not used for one or more of the following reasons: 1) the satellite has a larger
PSF and a much lower sensitivity than \chandra, allowing the cluster emission to
dominate over weak sources at that offset; 2) the data were taken during a 
series of \chandra\ non-detections; 3) the exposure was relatively short.

\subsection{The Transient}

\subsubsection{EUVE}

\euve\ pointed in the direction of A1795 seven times over 2.5 years. For all the
observations we analyzed the events recorded with the Deep Survey telescope 
\citep[DS;][]{bow91} with the Lexan/Boron filter in the 0.0404--0.2816
keV (67--178 \AA) energy range. Significant emission was detected only in a 
$\sim 70.8$ ksec observation on 1998 March 27. In an earlier, very long 
observation ($\sim 90$ ksec) on 1997 February 3 and in five shorter following 
pointings (up to $\sim 25$ ksec) in 1999, we were able to detect only emission
coinciding with the center of the galaxy cluster, and in particular with the 
galaxy B2 1346+26 (the \euve\ bright source J1348+26.5B). 

The cluster was the target of all the observations and, consequently, it was 
always on-axis, allowing us to consider the \euve\ PSF as undistorted (the DS
focal plane is curved, while the detector is flat). Since the distance of CXO 
J1348 from the central galaxy is $\sim 0.7$\arcmin\ and the angular 
resolution of \euve\ is $\sim 0.3$\arcmin, the emission from the two objects 
overlaps. 

To estimate the net flux associated with the transient we adopted the following 
procedure: We extracted the surface brightness profile (ct s$^{-1}$ 
arcmin$^{-2}$) 
for both of the long observations (in 1997 and in 1998) using the reprocessed 
images downloaded from the \heasarc\ archive and a series of annular regions 
centered at the cluster position. The annuli were 0.4\arcmin\ wide. We used the 
deadtime-corrected exposure time from the header of the FITS files. A 
comparison of the two profiles shows that the transient is significant within 
the first 1.5\arcmin\ from the cluster center. The counts associated with the 
cluster extend up to 4\arcmin. We extracted the net count rate from the inner 
1.5\arcmin\ region for both epochs and an annular region (4-5\arcmin\ as 
inner/outer radii) as background. To generate a response file, we downloaded the
effective area for the DS Lexar/Boron filter from the \euve\ handbook.

Using \verb+Xspec+ for the observation in 1997, we created a spectrum with
one bin covering the 0.0404--0.2816 keV range. Using the \verb+APEC+ thermal
model \citep[with $kT = 0.13$ keV and abundances 0.31 $Z_{\odot}$;][]{bon01}
to model the soft cluster emission in the \euve\ range, we found an observed 
flux of $(1.36\pm0.06) \times 10^{-12}$ \flux. We generated a spectrum also for
the 1998 observation using a combined model: the thermal model above and a 
black-body (\verb+BB+) component with a temperature of $kT = 0.09$ keV (see 
Sect.~2.1.2). Subtracting the cluster contribution from the total flux, 
$(5.43\pm0.12) \times 10^{-12}$ \flux, the observed flux from CXO J1348 is 
$(4.07\pm0.13) \times 10^{-12}$ \flux.
Using \verb+WebPIMMS+, the flux was extrapolated to the \chandra\ 0.5--7 
keV energy range. As a result we found that the absorbed and unabsorbed fluxes
are $(3.15\pm0.10)$ and $(3.41\pm0.10) \times 10^{-12}$ \flux, respectively,
at this time.

\subsubsection{Chandra}

\begin{figure} 
\center
\includegraphics[scale=0.8]{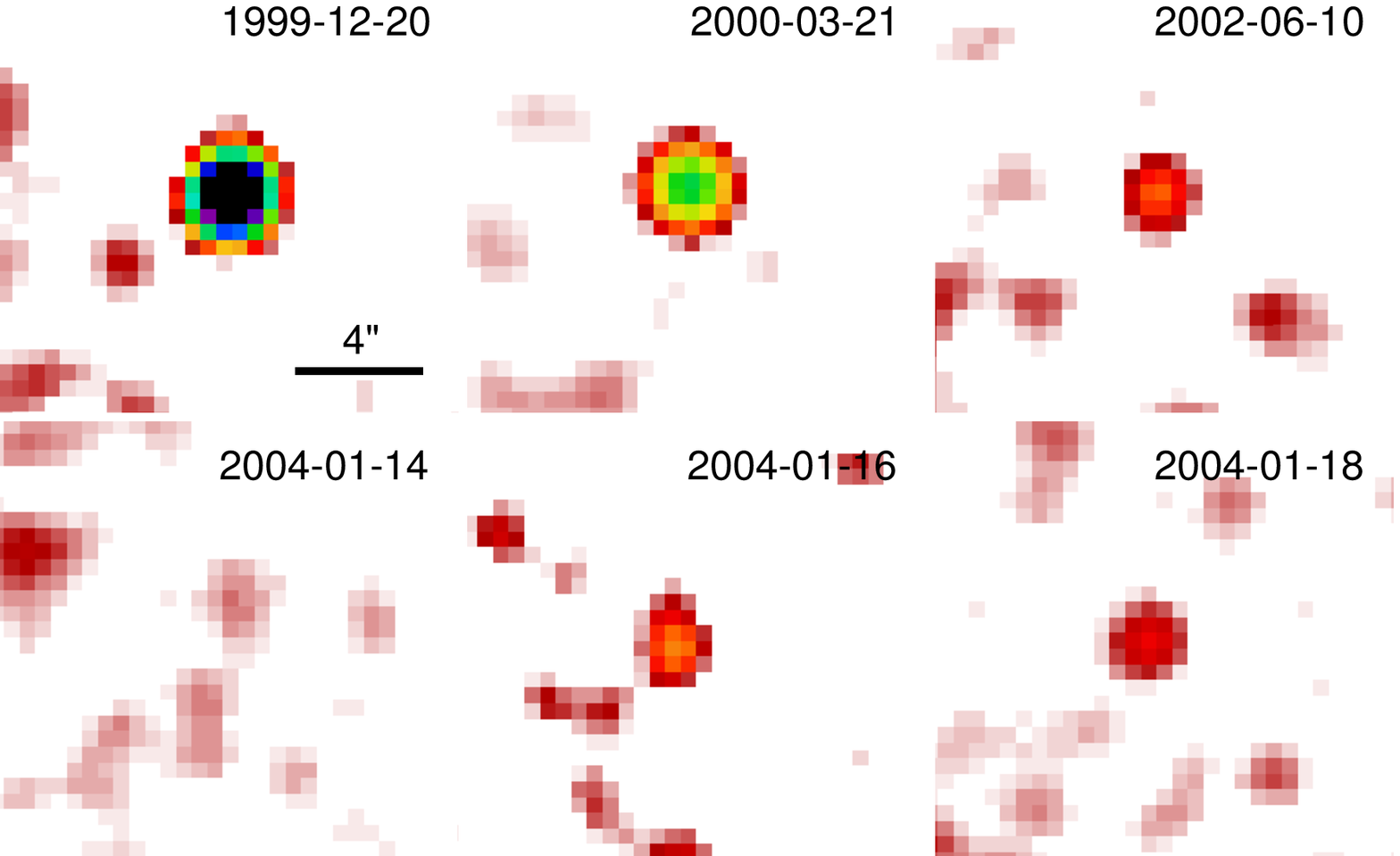}
\caption{
Field of view of the \chandra\ observations centered at the transient location.
The 0.3--8 keV images was smoothed with a 3-pixel Gaussian 
function in {\it ds9}. \label{chandrafield}
}
\end{figure}

A1795 is a familiar target for \chandra\ \citep{wei00}: except for the 
first 2 pointings (PI: 
Fabian), the cluster has been used as a calibration source. Up to March 2012, 
the telescope observed this field 43 times. Visual inspection of the images 
reveals that for only 26 pointings the position of CXO J1348 was within the 
boundaries of the ACIS (I or S) detectors. The list of useable observations
is given in Table~\ref{xray}. As mentioned before, only exposures up to 2004 
January 18 detect emission from CXO J1348 (see Figure~\ref{chandrafield} for a 
snapshot of all the \chandra\ observations from 1999 to 2004). Running 
\verb+wavdetect+ we found the following coordinates: $\alpha = 13^h 48^m 
49.87^s$; $\delta = +26^{\circ}$ 35\arcmin\ 57.6\arcsec, with a systematic error
of 0.6\arcsec. As explained in the \verb+CIAO 4.4+ science thread, after 
correcting the aspect files we merged the ACIS-I data obtained in 2005 March and
all the remaining ACIS-I and ACIS-S observations up to 2012. No significant 
excess was found at the transient position.

Before performing the data analysis of those observations with a detection, we
re-generated the event 2 files using the \verb+chandra_repro+ script. We 
checked for the presence of flaring activity in the ACIS background: 
only very short time intervals were excluded from the following analysis. We 
selected the larger energy range 0.3--8 keV to estimate the net count rate and
the source significance, while we limited the spectral 
analysis to the artifact-free 0.5--7 keV range. The spectral files for the 
source and the background as well as the response files were generated by the 
tool \verb+specextract+, using a circular source extraction region with radius 
of 2\arcsec\ for the 1999 observation, when the source was brightest, 
1.5\arcsec\ for the observation in 2000, and 1.0\arcsec\ for the remaining 
observations in 2002 and 2004. Smaller extraction regions were chosen in 
later epochs to reduce the contamination from the cluster X-ray light. A 
smoothed image shows that the cluster emission still has a high gradient at the 
position of CXO J1348. For this reason, we selected as background regions 2 
boxes along the isophotes of the smoothed emission,
positioned on the two sides of the transient. The boxes are 4\arcsec\ wide and
10\arcsec\ long. The source spectrum was grouped to a minimum number of 15 
counts per channel when the source was bright (in 1999 and 2000).

\begin{figure}
\center
\includegraphics[scale=0.34]{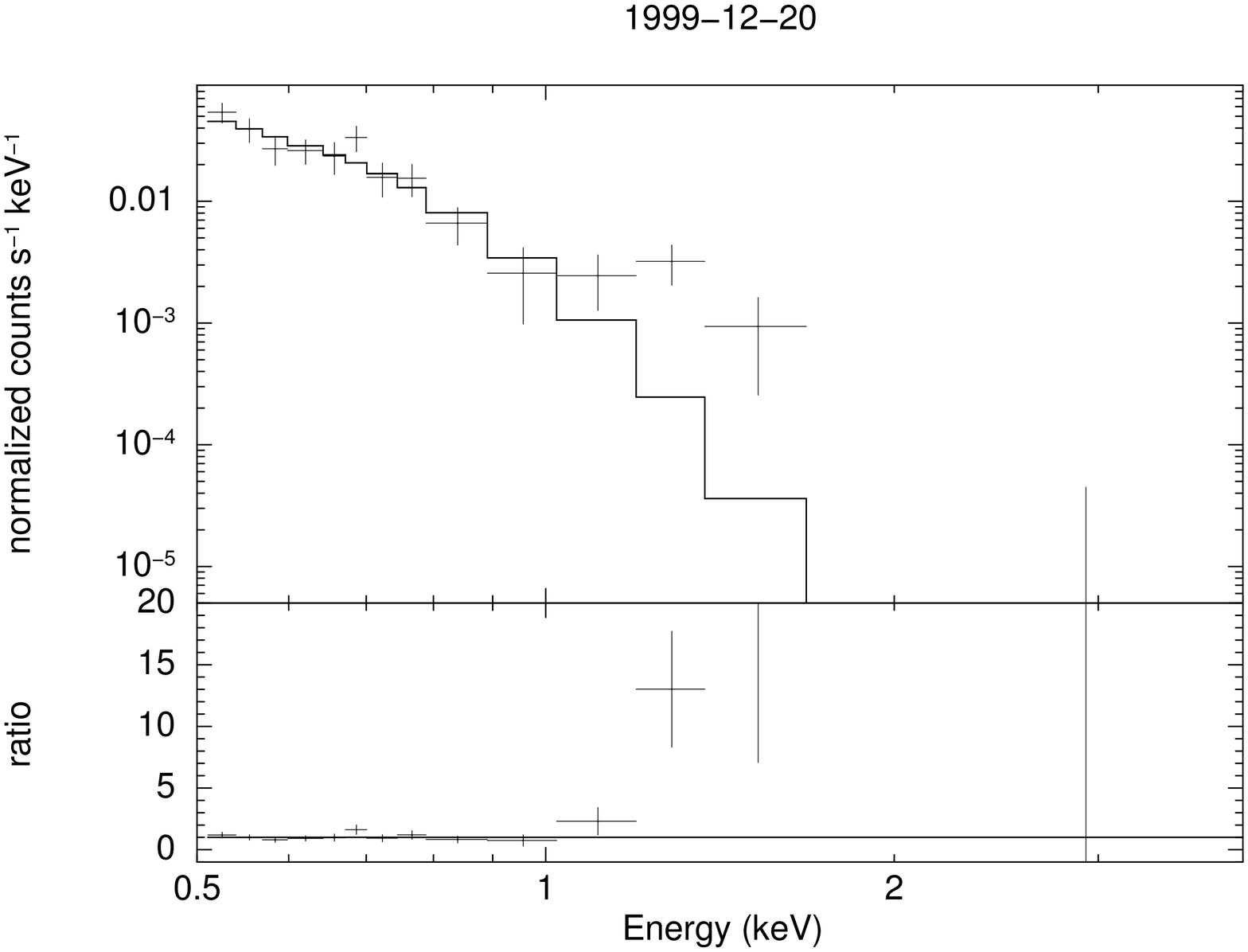}\includegraphics[scale=0.34]{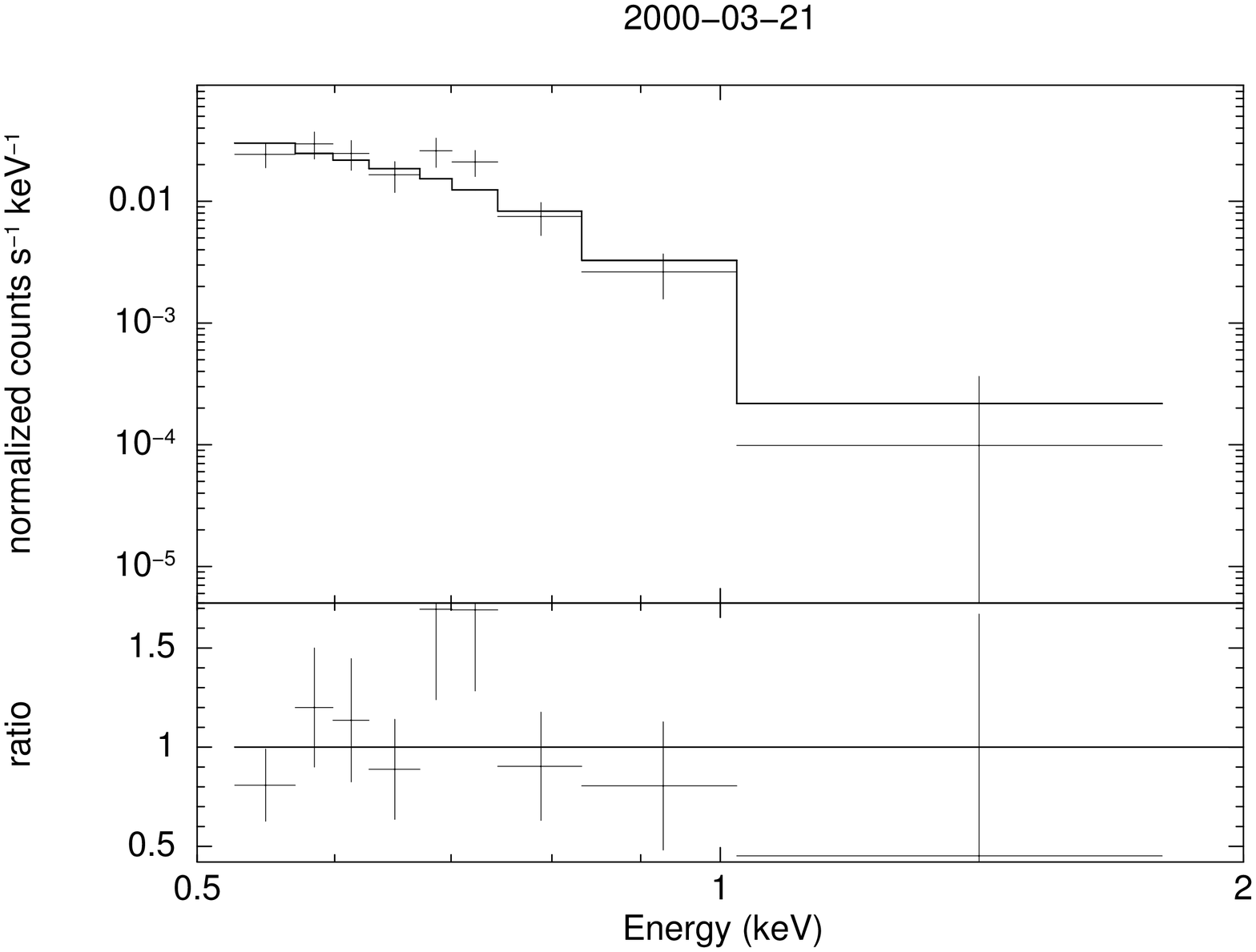}
\caption{
\chandra\ 0.5--7 keV spectra of the first two observations of CXO J1348. The 
observations taken on 1999 December 20 (left) and on 2000 March 21 (right) were 
fit using a blackbody model with $kT = 0.10$ and 0.09 keV, respectively. 
\label{chandraspectra}
}
\end{figure}

The 0.5--7 keV spectrum of CXO J1348 in 1999 can be adequately fitted 
($\chi^2 = 15.5$ for 12 degrees of freedom, d.o.f.) by a single \verb+BB+ model
with $kT = 0.10 \pm 0.01$ keV and absorption fixed at the Galactic level (left 
panel of Figure~\ref{chandraspectra}). The unabsorbed flux is $(2.8 \pm 0.4) 
\times 10^{-14}$ \flux. There is a hint of an excess above 1 keV and we added a 
second \verb+BB+ component to improve the fit. We found that the new fit 
($\chi^2 = 10.3$ for 10 d.o.f.) is obtained with $kT_{\rm 1} = 0.08 \pm 0.02$ keV
and $kT_{\rm 2} = 0.25 \pm 0.19$ keV. The second component is not statistically 
significant as the probability obtained with the $F-$test is only 13\%. Using 
more complicated spectral models (\verb+diskbb+ or \verb+diskpbb+) does not
improve the fit, since they still do not compensate for the excess above 1 keV. 
Alternatively, a good fit can be obtained using a bremsstrahlung model. We 
found a temperature of $0.19 \pm 0.03$ keV ($\chi^2 = 13.6$ for 12 d.o.f.) and 
an absorbed flux of $(2.6 \pm 0.4) \times 10^{-14}$ \flux. A fit with a
 power-law with a photon index of $\Gamma = 4.35 \pm 0.27$ is also a reasonable 
representation of the X-ray spectrum ($\chi^2 = 11.6$ for 12 d.o.f.). Such a 
steep spectrum typically mimics thermal emission over the limited \chandra\ 
bandpass. Similar results have been found using an un-grouped spectrum and 
the $C-$statistic.

The observation obtained in 2000 was well fit ($\chi^2 = 7.8$ for 7 d.o.f.) by
an absorbed \verb+BB+ model (right panel of Figure~\ref{chandraspectra}). The 
temperature is $kT = 0.09 \pm 0.01$ keV and the unabsorbed flux is $(2.1 \pm 
0.4) \times 10^{-14}$ \flux. No emission associated with the transient is 
observed above 2 keV. The use of a bremsstrahlung model does not fit the 
spectrum better: although the temperature ($kT = 0.16 \pm 0.03$ keV) is very 
similar to the 1999 observation, the residuals are higher ($\chi^2 = 9.8$ for 7 
d.o.f.).

In 2002 the source faded significantly and only 27 net counts in the 0.3--8 keV 
range are possibly associated with CXO J1348. The detection significance is 
4.1$\sigma$ (that increases to 5.3$\sigma$ in the 0.3--2 keV range). A spectral
analysis using the $C-$statistic still shows the presence of the thermal
\verb+BB+ component ($kT = 0.07\pm0.03$ keV) but at lower intensity 
(2$^{+2}_{-1} \times 10^{-15}$ \flux\ in the 0.5--7 keV range). 

In all these trials we allowed an additional absorption at the cluster
distance to vary, but its value was either small/unconstrained or did not 
improve the fit.

In 2004, the source was observed three times in a four day time span and it 
showed 
some signs of flaring activity: on January 14, the source was not significantly
detected on the ACIS-S detector (0.9$\sigma$, with an upper limit on the count 
rate of $< 4.7 \times 10^{-4}$ in the 0.3--8 keV range), but it re-appeared on 
January 16 with a detection significance of 3.5$\sigma$ [ACIS-S net count 
rate $(1.4 \pm 0.4) \times 10^{-3}$ ct s$^{-1}$]. The transient was visible also
on January 18 with a significance of 3.5$\sigma$ (ACIS-I net count rate $(1.3 
\pm 0.3) \times 10^{-3}$ ct s$^{-1}$). Due to the extremely poor statistics, 
no spectral analysis was performed. The unabsorbed flux in the 0.5--7 keV range
are $(4 \pm 1) \times 10^{-15}$ \flux\ and $(1.5 \pm 0.5) \times 10^{-14}$ \flux,
using the \chandra\ \verb+PIMMS+ tool \footnote{Although the net count rates in
2002 and 2004 are almost identical, due to the evolution over time of the 
\chandra\ response matrices, PIMMS predicts higher fluxes for the ACIS-I 
detector.} and assuming a \verb+BB+ model with a temperature of 0.09 keV 
(similar to the values found in the 1999--2002 period). For the observation on 
January 14, we estimated a 3$\sigma$ upper limit of $< 1.5 \times 10^{-15}$ 
\flux.

Visual inspection of all the pointings starting in 2005 does not reveal the 
presence of the transient. We estimated an upper limit from the first pointing 
obtained on 2005 March 20 because the most sensitive detector ACIS-S was used. 
Since the cluster is the dominant source of emission, the upper limit remains 
high at $< 1.4 \times 10^{-15}$ \flux.

\subsubsection{VLA}

The field of A1795 has been extensively observed with the NRAO\footnote{The 
National Radio Astronomy Observatory is a facility of the National Science 
Foundation operated under cooperative agreement by Associated Universities, 
Inc.} Very Large Array (VLA) due to interest in its central bright radio 
galaxy, 1356+268 \citep[e.g.,][]{ge93}. Considering the time span after the 
detection of the transient X-ray source, we selected and analyzed archival VLA 
data of the field consisting of observations obtained at three epochs from 2000
October to 2005 October in the $\sim 5-8$ GHz range. All on-source exposures 
were single snapshots lasting $1-10$ min, using various VLA configurations. The
data were calibrated in AIPS using standard procedures and self-calibration and 
imaging were performed with DIFMAP \citep{she94}. 
No significant radio emission was detected at the position of CXO J1348 
in any of the VLA observations with point source limits ranging from 
$<0.10$ to $<0.32$ mJy (see Table~\ref{radio}).

\subsection{The Host Galaxy}

\begin{figure}
\center
\includegraphics[scale=0.51]{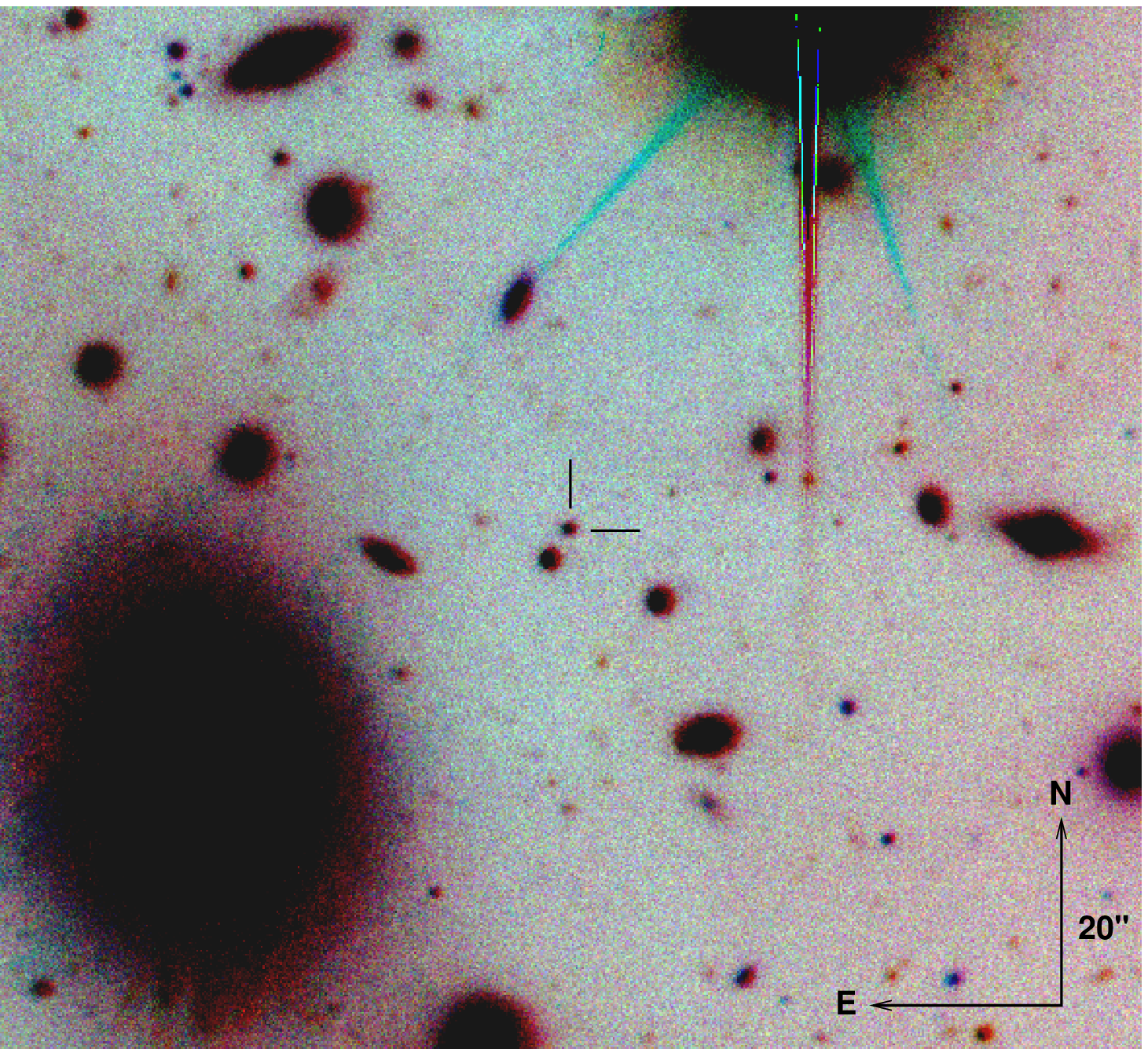}\includegraphics[scale=0.33]{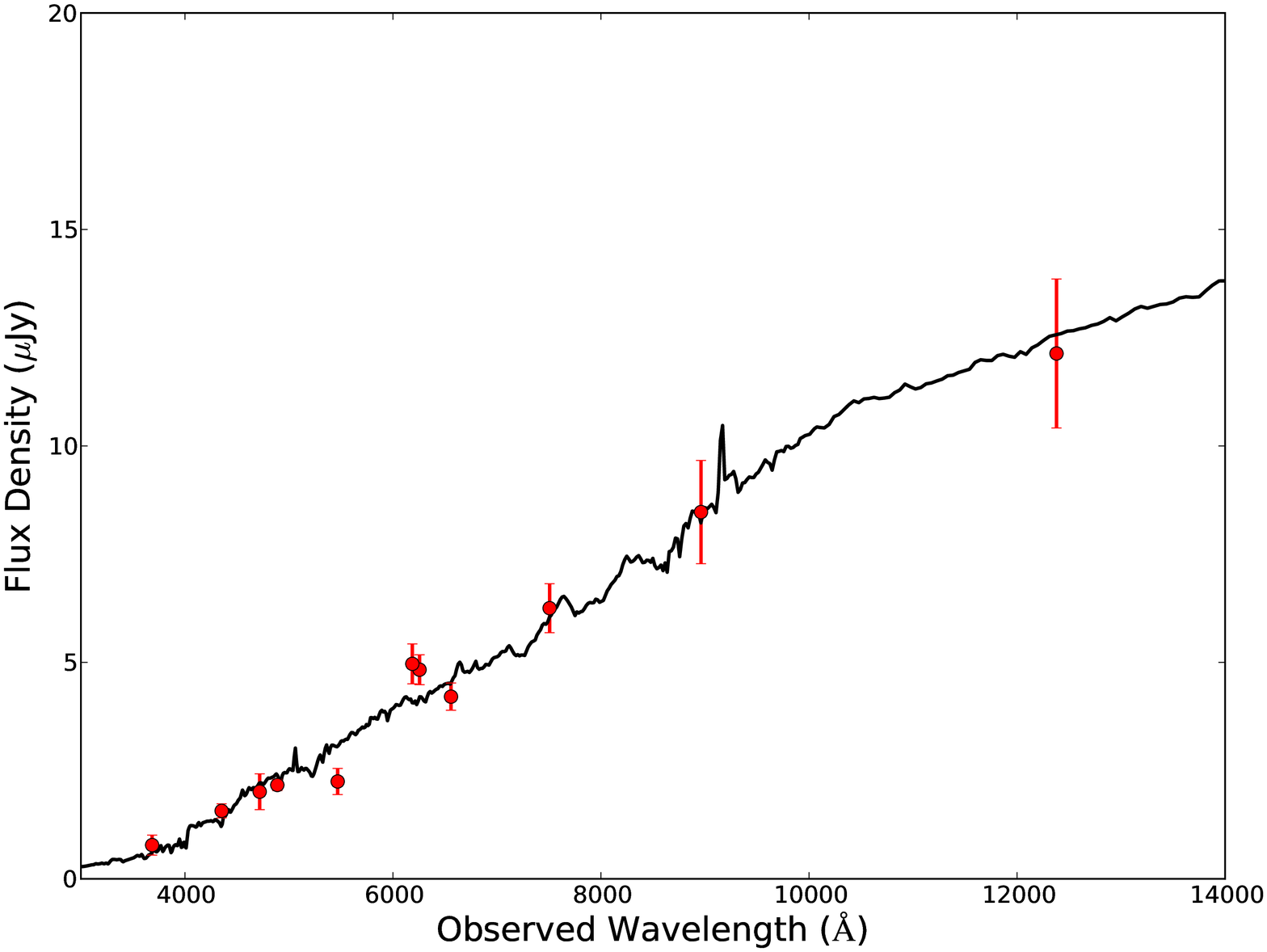}
\caption{
{\bf{Left}}: False color image combining CHFT $g'$ and $r'$ and NOT $i'$ 
bands. The host galaxy of CXO J1348 is marked at the center of the image.
{\bf{Right}}: SED obtained by combining all the new and archival optical and 
near-IR photometric measurements. The black line indicates 
the best fit found by running the photo-redshift code {\it EaZy}, using an 
evolved stellar population model at redshift $z = 0.13^{+0.18}_{-0.05}$. 
\label{vltfield}
}
\end{figure}

A1795 has been observed extensively over the years in the optical by ground- and
space-based telescopes. Coincident with the X-ray transient, we find a faint,
resolved source which we shall assume to be the host galaxy of CXO J1348 (see 
Figure~\ref{vltfield}, left panel). Here we describe both new and archival 
observations of this galaxy, with the aim of constraining its distance and other
basic properties (BH mass, nuclear activity, etc.).

The photometry of ground-based telescopes was performed using the {\tt DAOPHOT} 
APPHOT photometry package in {\tt IRAF}. Calibration was performed using field 
stars with reported fluxes in both the Two Micron All Sky Survey 
\citep[2MASS;][]{skr06} and the SDSS Data Release 9 Catalogue \citep{ahn12}. The
optical and infrared photometry of new and archival data is summarized in 
Table~\ref{optical}. The values have all been corrected for Galactic foreground
extinction, assuming $E(B-V) = 0.012$ \citep{sch11} and a Milky Way extinction
law with $R_{\rm V} = 3.1$ \citep{car89}.

\subsubsection{New Observations}

\noindent {\bf Observatorio Astrono\'mico Nacional/San Pedro M\'artir (OAN/SPM)
Johnson Telescope:} Data were obtained with the multi-channel Reionization And 
Transients InfraRed camera \citep[RATIR;][]{but12,wat12} mounted on the 1.5-m 
OAN/SPM telescope in Baja California, M\'exico.  On 2013 February 12 and 19, we
took a series of 60 sec exposures with dithering between them in various filters
(the number of exposures are indicate in parenthesis): \fg\ (80), \fr\ (80), 
\fii\ (160), \fz\ (120), \fJ\ (120), and \fH\ (22). Given
the small galaxy size, a sky frame was created from a median stack of all the 
images in each filter. Flat-field frames consist of evening sky exposures.
Due to lack of a cold shutter in RATIR's design, IR darks are not available. 
Laboratory testing, however, confirms that dark current is negligible in both 
IR detectors \citep{fox12}. The photometric images were reduced and co-added 
using standard CCD and IR processing techniques in IDL and Python.

\vspace{2 mm}
\noindent {\bf Nordic Optical Telescope (NOT):} 
We obtained three images in the \fB\ and \fii\ filters on 2012 March 20 with the
ALFOSC camera and five images in the \fU\ filter on 2013 March 14 with the MOSCA
camera mounted on NOT \citep{kar93}. Exposure times were 500, 300, and 600
s, respectively. On both occasions, the sky conditions were photometric, however
the seeing was variable. The frames were reduced and co-added using standard 
IRAF procedures (de-biasing, flatfield correction). 

\vspace{2 mm}
\noindent {\bf Large Binocular Telescope (LBT):} 
On 2013 April 2 we obtained an optical spectrum
with the Multi-Object Double Spectrograph (MODS1) instrument at the focus of the
two 8-meter mirrors of the LBT \citep{hil00} using a 1.0\arcsec\ wide slit
and the G400L and G670L grisms for the blue and red channels, respectively, 
covering the 3200-5800 \AA\ and 5800-10000 \AA\ range. The Clear filter was used
for both grisms. The spectral resolution is $\lambda/\delta\lambda \sim 200-400$
km s$^{-1}$, depending on the spectral region. Since MODS1 does not have an 
atmospheric dispersion corrector, the slit was oriented along the mean 
parallactic angle (PA =70\degr). Conditions were clear with an average seeing 
always better than 1.5\arcsec\ FWHM and the observations were done in a 
sequence of four 1800 s exposures for a total integration time of 2 hr. The data
were reduced by the LBT data center. 
Since the optical host is a dim object, we were able to detect only a weak 
continuum emission, with signal-to-noise ratio (SNR) $\approx 3$ per pixel 
($\approx 8$ per resolution element) in the range from 6000--8000\,\AA.  Due to 
the red galaxy color, the SNR decreases as a function of wavelength and no 
significant signal is detected below $\approx 4000$\,\AA.  No significant 
features are observed over the range from 4000--9500\,\AA, neither in absorption
nor emission.  Specifically, for the region from $\approx 6000$--8000\,\AA, we 
limit the flux from any emission line to be $f \lesssim 10^{-17}$\flux\ 
(assuming a line width of several hundred km s$^{-1}$, corresponding to our 
instrumental resolution).

\subsubsection{Archival Images}

\noindent {\bf Hubble Space Telescope (HST)}: 
Although A1795 has been the target of many \hst\ pointings, due to the very 
small field of view CXO J1348 fell on a detector only in a single set of
observations. On 1999 April 11 the telescope observed the source position 
using the Wide Field Planetary Camera 2 \citep{hol95} for 300 sec, both 
with the F555W and the F814W filters. Unfortunately, at that time the drizzling 
technique to remove cosmic rays was not adopted and single images (per 
filter) were taken. An optical counterpart at the position of the X-ray 
transient is detected on the WF2 chip in both filters. A visual inspection of 
the processed images revealed the presence of cosmic rays very close to the 
counterpart in the F814W image. The photometry in this filter is, thus, 
unreliable. The estimated flux in the F555W filter is corrected for the finite 
aperture using Table~2b in \citet{hol95}.

We also check the morphology of the optical source by comparing it with an 
artificial PSF. We used the web interface of the PSF modeling tool 
\verb+Tiny Tim+ \citep{kri11} to generate a PSF located at the same position on
the WF2 chip of the F555W filter, and we assumed a spectrum described by a power
law with index $-1$ (changing the spectral slope does not alter the shape of the
PSF significantly). We selected the F555W filter because there were no 
identifiable cosmic rays close to the source. Since the object is very dim and 
the PSF is under-sampled, we smoothed the image using a Gaussian function with 
a 2 pixel kernel radius. We extracted a brightness profile along the East-West 
direction (the diffraction spikes do not contribute since they are tilted by 
10\degr). The profile of the source was compared with that of the generated PSF,
and smoothed with
the same kernel function. The Gaussian function that describes the artificial 
PSF has a FWHM of 0.28\arcsec, while the source brightness profile has a larger
width (FWHM = 0.40\arcsec). This indicates that the source is spatially 
resolved, suggesting a galaxy-like morphology. A comparison between the images 
in the two filters shows that there is some additional emission above the PSF 
and 3$\sigma$ above the local background located at PA = 215\degr\ and 
with extension of 0.6\arcsec. The source measures 1.05\arcsec\ along this angle 
and 0.7\arcsec\ perpendicularly. The short exposure of the two images did not 
allow us to speculate on the nature of this feature. 

\vspace{2 mm}
\noindent {\bf Very Large Telescope (VLT)} and {\bf Canada-France-Hawaii Telescope (CFHT)}:
The optical counterpart at the position of CXO J1348 has been observed with the
FORS1 camera on the VLT \citep{nic97} on 2002 June 08 and July 19, and the
MegaPrime/MegaCam on the CFHT \citep{bou03} on 2008 August 5 and 
2009 June 25. Frames from the two telescopes have been corrected by means of 
bias or dark frames and response was normalized by means of flat-field frames. 
Given the lack of variability due to the temporal proximity of the two sets of
VLT data, we decided to sum the observations to increase the signal-to-noise 
ratio, in particular in the $U$ filter where the source was barely detected.

\vspace{2 mm}
\noindent {\bf Isaac Newton Telescope (INT)}: 
Querying the online catalogs at \heasarc\, we found that photometric 
measurements for an object consistent with the \chandra\ source were already 
available. The optical source was observed as part of the WIde-field Nearby 
Galaxy-cluster Survey \citep[WINGS;][]{var09} between 2000 and 2001. The 
automatic software that run the analysis in the \fB\ and \fV\ filters determined
that the object (WINGS J134849.88+263557.5) can be classified as a galaxy. The
galaxy cluster has been extensively observed through the years, from 1992 to
2010, using the Wide Field Camera on the INT \citep{lew00}. While the 
images in the \fr\ and \fii\ filters suffer from bad fringing, the observations 
in the \fU, \fB, and \fV\ can be used to extract valuable photometry. The source
does not show any sign of variability. In Table~\ref{optical} we report the 
values obtained in the 2010 May campaign, a period not covered by other 
ground-based telescopes.

\vspace{2 mm}
\noindent {\bf Spitzer}:
Only one imaging data set covering the location of CXO J1348 is available in the
\spitzer\ archive. It was taken on 2010 August 8 with the IRAC instrument 
\citep{faz04} in both the 3.6 $\mu$m and 4.5 $\mu$m bandpasses. A visual 
inspection reveals that only in the 3.6 $\mu$m (a 380 s exposure) mosaic image 
there is a source at the X-ray transient position detected at the 2$\sigma$ 
level, while in the 4.5 $\mu$m bandpass the significance is 1$\sigma$. 
Unfortunately, the relatively short exposures and the contamination from the 
galaxy at the center of the cluster do not allow us to extract a reliable and 
meaningful photometric measurement: using circular regions for both the source 
and the background with a 2 pixel radius and applying the aperture corrections 
listed in Section 4.10 of the IRAC instrument handbook, we obtained flux 
densities of 12.8$\pm$7.7 $\mu$Jy and 3.4$\pm$3.0 $\mu$Jy in the 3.6 $\mu$m and
4.5 $\mu$m bandpasses, respectively. Due to the large errors, we do not use 
these values in the analysis.

\vspace{2 mm}
\noindent {\bf Swift Ultra-Violet/Optical Telescope (UVOT)}:
A1795 was observed with the UVOT \citep{rom05} in the \fW\ filter on-board the 
\swift\ satellite on 2005 
November 12. We used \verb+uvotimsum+ to combine the total of $\sim$ 24 ksec 
exposure obtained over three orbits. Running \verb+uvotsource+ we found that 
no significant emission above the background was observed at the X-ray position 
with an upper limit, $m_{\rm UVW1}<22.4$ (in the Vega system).

\subsubsection{Characteristics of the Host Galaxy}

Since no spectroscopic redshift of the host galaxy is available, we combined 
photometry from archival VLT, CFHT, INT and \hst\ data with new observations 
from RATIR and NOT to produce a detailed SED of the host galaxy (right panel 
of Figure~\ref{vltfield}).

We fitted the SED using the \verb+EaZy+ code
\citep{bra08} and models from \citet{bru03}: We considered different kinds of 
galaxy spectral templates (elliptical, early and late spirals, irregular and 
starburst galaxies), set at different ages and with various metallicities and 
star formation rates. Although many 
templates provide a reasonable fit of the putative host galaxy SED, the best fit
is obtained by either an elliptical or S0 template, both dominated by an old, 
evolved stellar population, located at a redshift of $z = 0.13^{+0.18}_{-0.05}$ 
(68\% confidence level). The derived photometric redshift is compatible 
with the average value of the cluster redshift ($z=0.062476$) and the dispersion
of the velocities of the cluster brightest galaxies, whose redshifts range from 
0.054 and 0.068 \citep{smi04}.

The inferred de-reddened magnitude at the cluster redshift is $M_{\rm B}\sim 
-13.8$ ($M_{\rm R}\sim -15.1$) and the estimated scale is 1.188 kpc arcsec$^{-1}$
\citep{wri06}. Based on the result of the spatial analysis of the \hst\ data, 
this corresponds to a minimum radial extension of 0.4 kpc (and up to 0.7 kpc 
along the putative feature observed to the south-west of the source). Thus, both
the source brightness and its radial extent suggest that we are observing either
a compact elliptical galaxy or a spiral galaxy with a small bulge and even 
dimmer spiral arms. Both scenarios are supported by the reasonable fit obtained 
using an elliptical or S0 template for the estimate of the photo-redshift.

The uncertainties on the photometric redshift are somewhat large in the upper 
end side: the 68\% error puts the host galaxy at $z = 0.31$, an increase by a 
factor of 5.8 in the luminosity distance and of 3.8 in the angular scale. Thus,
the source would be 3.8 magnitude brighter and 1.5 kpc in size. 

The assumption that the host galaxy belongs to the cluster A1795 is supported
by its projected location, very close to the main galaxy of the cluster, and by 
the richness of the cluster. Querying the SDSS catalog for an area with a radius
of 20\arcmin, where previous works \citep[see, e.g.,][]{oeg01} have found the 
majority of the cluster components, we find that more than 150 galaxies 
(corresponding to 3 out of 4 galaxies with a spectroscopic measurement) have a 
redshift compatible with the cluster. We can not rule out entirely
the possibility that the host galaxy belongs to a more distant group or is a 
background object. Of the remaining galaxies with an estimate of their
distance, 20 sources have a redshift of $z = 0.11$, indicating that another
cluster may be located behind A1795, $\sim$15 objects have redshift more evenly 
distributed in the ranges 0.15--0.20 and 0.24--0.31, while a few other sources 
are located at cosmological distances.

\section{Origin of the UV/X-ray Transient}

The spatial analysis of the \hst\ and ground-based data shows that the optical 
counterpart at the position of the UV/X-ray transient is not a point source. 
This excludes any local origin (e.g., explosion in a classical or a recurrent 
nova system, X-ray burst on the surface of a neutron star, etc.) and leads to
the conclusion that we are observing an extragalactic object, i.e., a distant
host galaxy. 

The probability of chance alignment of the optical source with a \chandra\ 
position is very low. Following \citet{blo02} and \citet{per12} an estimate of 
the probability of chance association $P$ can be expressed as

\begin{equation}
P = 1 - exp^{- A \rho}.
\end{equation}

\noindent Here, $A$ is the area on the sky encompassing the X-ray and the 
optical sources, while $\rho$ is the sky density of objects of equal or greater 
brightness. Very conservately we choose a circle with a radius of 2\arcsec\ as 
the area, that 
corresponds to the largest X-ray flux extraction region. There are other factors
that might contribute to the size of that circle, like the \chandra\ astrometry 
error (typically of the order of 0.6\arcsec) or the \hst\ PSF of the object 
(0.4\arcsec, see above), but their contribution is not significant when compared
to the size of the extraction region. We queried the SDSS catalog to have a list
of sources brighter than the host galaxy 23.14 mag in the {\it g'} filter in a
circle with a radius of 15\arcmin\ and centered at the transient position. Since
we found $\sim 2000$ objects, the chance probability is, thus, $<0.01$. Assuming
the quadratic sum of \chandra\ and \hst\ astrometric errors only, the 
probability is $\sim 0.001$.

A similar argument can be done to estimate the chance probability that the
transients observed by \euve\ and \chandra\ are actually the same object. This
assumption is hampered by the fact that there is a 21-month gap between the 1998
March \euve\ and the 1999 December \chandra\ observations, which includes 
multiple \euve\ non-detections. The difference with the previous approach is
that there are no sources in both fields brighter than the transient. The only 
other point-like object detected by \euve\ is EUVE J1348+26.5A, a Seyfert 1 
galaxy located at 5.8\arcmin\ from the transient \citep[from the second 
\euve\ right angle program catalog in][]{chr99}. This object is also the second 
brightest point source in all the combined \chandra\ fields of A1795, after the 
transient. Assuming an extraction region of 1.5\arcmin\ (see the \euve\ analysis
in Sect.~2.1.1), the estimated chance probability is $\sim 0.065$. 
This does not include the likelihood or another (unrelated) high-amplitude
flare in the field, however.  While difficult to quantify, we can use results 
from the \galex\ Time Domain Survey \citep{gez13} and the \xmm\ Slew Survey 
\citep{sax08} to estimate the probability of an unrelated transient source in 
our \euve\ images.  The sky density of highly variable ($\Delta$mag $\ge 2$ in 
the \galex\ NUV filter) M dwarf flares (the dominant class of such dramatic 
variability) is $\sim 5$\,deg${^2}$\,yr$^{-1}$ \citep{gez13}.  Given the 
duration of the \euve\ exposure (70.8\,ks) and the astrometric uncertainty 
(1.5\arcmin), the likelihood of chance detection of an unrelated source is only
$\sim 2 \times 10^{-5}$.  Similarly, \citet{kan13} calculate the rate of X--ray
transients of comparable brightness to CXO\,J1348 is $3 \times 
10^{-3}$\,deg$^{-2}$.  The implied probability of detecting an unrelated source 
is again very small, $\sim 6 \times 10^{-6}$.  We conclude the \euve\ detection 
is highly likely associated with CXO\,J1438.

Here we consider four possible classes of extragalactic transients known to
produce the observed degree of X-ray variability: a GRB, a supernova, a sudden 
outburst from an active galactic nucleus (AGN), or a TDE.

\subsection{GRB / Supernova Scenario}
The most compelling lines of evidences against the GRB/supernova scenario are: 
1) the length of the UV/X-ray light curve; 2) the short term variability at 
late time; 3) the shape of the X-ray spectrum. Below we discuss each in more 
detail.

CXO J1348 is detected for approximately six years, a period over which the 
intensity decays by 3 orders of magnitude (from the \euve\ detection in 1998 
and the \chandra\ upper limit in early 2004 there is a factor of $\sim$2300). 
\swift\ monitoring of hundreds of GRB afterglows \citep{dav12}
shows similar decrease in brightness, but they do not last as long: the longest 
light curves (e.g., GRB 060729) span only a time frame of a few months before
reaching a flux level of $\sim10^{-14}$ \flux\ (a value similar to
the last detection of CXO J1348 by \chandra\ in 2004). 

On the contrary, supernovae can be detected in the X-rays for decades 
\citep{dwa12} but the light curve does not decay as much and, in some cases, it
rises over time due to the interaction of the shock waves with the 
circumstellar medium. 
As mentioned in Sect.~2, the \chandra\ data taken five years after the first 
\euve\ observation show dramatic variability over a time of four days. This kind
of late time variation is strong evidence against the supernova scenario, since 
a typical X-ray light curve is not expected to show fast flaring activity many 
years after the original explosion.

The result of the \chandra\ spectral analysis can be used to rule out the GRB 
afterglow scenario. Their X-ray spectra are fitted by a power-law model with a
flat spectral slope \citep[$\Gamma \simlt 2$,][]{but07}, while a small fraction 
(10\%) of the afterglows have a prominent residuals near 1 keV that can be 
fitted with an additional blackbody model. On the contrary, the first 
observation with \chandra\ of CXO J1348 was fit with a much steeper power-law 
($\Gamma \sim 4.3$) or by simple thermal model, as are all the following 
spectra.

Furthermore, the host shows no evidence for star formation. Long GRBs are found
exclusively in star-forming galaxies \citep[e.g.,][]{fru06,sav09} and 
core-collapse supernovae are the only ones for which X-ray emission has been 
detected.

\subsection{AGN Scenario}

The lack of emission lines in the optical LBT spectrum might be intrinsic,
as seen in blazars, a sub-class of radio-loud AGN \citep{urr95}, or 
because the lines were too dim to be detected. Optical broad or narrow emission
lines, like H$\alpha$, H$\beta$, etc. are observed in a large variety of
AGN, all of them showing some X-ray variability. At the redshift of the cluster,
our limits on any H$\alpha$ emission correspond to a 
luminosity of $L_{{\rm H}\alpha} < 8 \times 10^{37}$\lum.  This value is several 
orders of magnitude less than the H$\alpha$ luminosities derived from 
\citet{gre07} for active galaxies with low-mass BHs.  

\subsubsection{Radio-loud AGN / Blazars}

The radio emission observed in these objects is produced by accelerated 
particles that travel in collimated, relativistic jets and emit synchrotron 
radiation.  As mentioned earlier, the object is not found in VLA archival images
down to a 3$\sigma$ flux limit of $< 0.1$ mJy. At a distance of $z \sim 0.07$, 
the radio luminosity is $L_{\rm 8.5 GHz} < 10^{38}$ \lum. 
This makes the putative AGN a radio-dim (or even a radio-quiet) source. Indeed, 
observations of the Hubble deep fields (HDFs) at 8.5 GHz show that all the 
radio sources have fluxes below the above upper limit and they are not 
identified with quasars but rather with star-forming galaxies, bright field 
elliptical and late-type galaxies with evidence of nuclear activity (Seyferts) 
and field spiral galaxies \citep{ric98}. The upper limit
excludes not only radio-loud galaxies but also radio-loud quasars (blazars): 
their radio luminosities start 2 orders of magnitude higher \citep{don01,gio12} 
and the broadband radio-to-optical spectral index ($\alpha_{\rm ro}$) between 
8.46 GHz and 5500 \AA\ is higher than the upper limit of 0.28 found for CXO 
J1348, a value that places the putative AGN in the radio-quiet regime.

Furthermore, blazars typically occur in very massive, luminous galaxies 
\citep[see, e.g.,][]{urr00} and there is some evidence that they might be
considered quasi-standard candles, with optical absolute magnitude greater than
$-24$, many orders of magnitude brighter than what is observed in this host 
galaxy.

\subsubsection{Radio-quiet AGN / Seyfert}

Among the radio-quiet AGN, like quasars, Seyfert galaxies, and low-ionization 
nuclear emission-line regions, narrow-line Seyfert 1 (NLSy1) galaxies are the 
most variable, although the changes are not as dramatic as in CXO J1348. In 
NLSy1 galaxies, the variability can be explained as due to changes in the 
properties of the absorber surrounding the inner BH such as column density, 
ionization parameter, and covering factor \citep[e.g.,][]{gru12}. The highest 
variability was seen in WPVS 007 \citep{gru95} with a decrease in flux by a 
factor of 400 in the soft band in a three year timespan, while fast variability 
has been observed in NGC 4051 \citep{mch04} with a change of the X-ray flux by 
a factor of 10 in a few days. This could explain changes seen in \chandra\
data alone, but not if we assume that the \euve\ transient is the same source.

One difference between CXO J1348 and the radio-quiet AGN is the shape of the
X-ray spectrum: radio-quiet AGN have their spectra described by a power-law with
spectral index in the range 1.5--3.5, irrespective if the source harbors a SMBH 
or an IMBH, and only the more luminous (and more massive) objects show the 
presence of the soft excess with temperature above 0.1 keV 
\citep{bra97,gru95,por04,min09,pia10,don12}. The NLSy1 galaxies do show a steep 
spectrum in the soft band (some of the \rosat\ spectra have slopes above 4), 
but they are detected at higher energies as well: the \asca\ spectra can be 
fitted in first approximation with a combination of a power-law model and a 
thermal blackbody component with temperature in the range 0.1-0.2 keV 
\citep{lei99}. On the contrary, the spectrum of CXO J1348 does not show any 
emission above 3 keV and, consequently, no additional power-law model is needed.

At optical wavelengths, \citet{zho06} found that in a sample of $\sim$2000 
optically selected NLSy1 in the SDSS, their absolute magnitude in the $g'$ 
filter is in the range $-18 \simlt M_{\rm g'} \simlt -26$, while the host galaxy 
of CXO J1348 is much dimmer ($M_{\rm g'} \sim -14.1$).

\subsection{TDE Scenario}

\begin{figure}
\center
\includegraphics[scale=0.6]{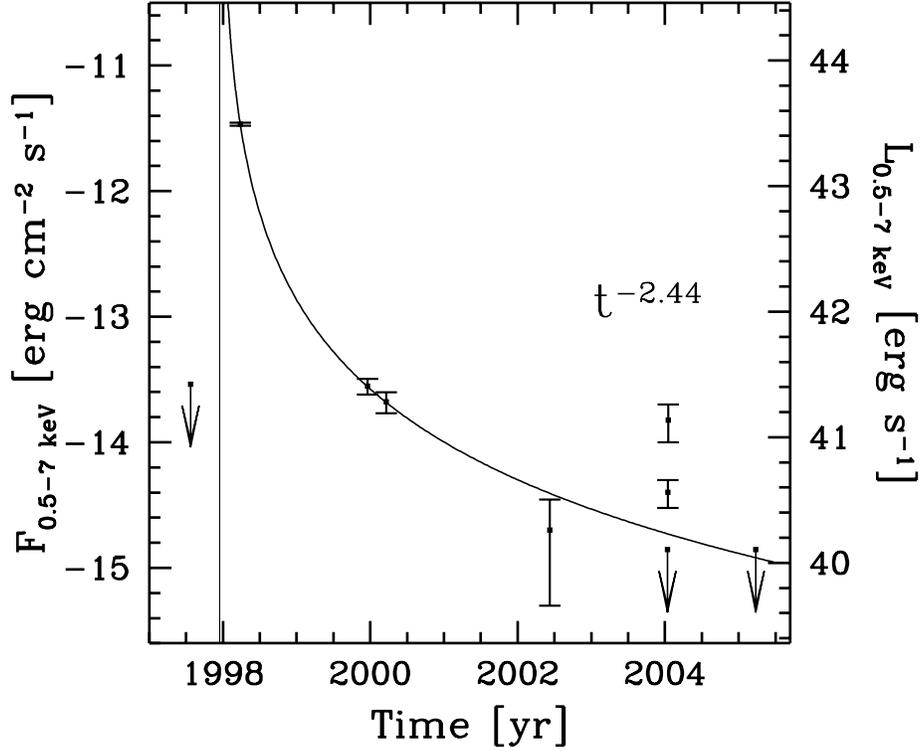}   
\caption{
Light curve for the UV/X-ray flare. The first point is an upper limit from 
\rosat, while the first detection is from \euve, and converted to the \chandra\ 
0.5--7 keV energy range assuming a black-body model with temperature of $0.09$ 
keV, as seen in following observations by \chandra. The last point is an upper 
limit from the ACIS-S observation on 2005 March 20. \label{lightcurve}
}
\end{figure}

The analysis of the combined \chandra\ and \euve\ light curve shows that the
brightness decay in the 0.5--7 keV range is consistent with a power-law decline
in time, characterized by $a \times [(t-t_{\rm D})/(t_{\rm 0}-t_{\rm D})]^{-n}$ 
law, where $a$ is the normalization and $t_{\rm D}$ and $t_{\rm 0}$ are the time 
of the disruption and of the flare's peak, respectively. Leaving all parameters
free to vary, we performed a least-squares fit of the decay (reduced $\chi^2 = 
3.1$ for 5 d.o.f.) and we found the 
following set of parameters: $a = (1.51 \pm 0.75) \times 10^{-13}$, $n = 2.44 
\pm 0.40$, and $t_{\rm D} = 1997.957 \pm 0.115$ (corresponding to 1997 December 
17 with an uncertainty of $\sim$42 days). While the value of $t_{\rm 0}$ 
remains unconstrained, $t_{\rm D}$ is consistent with the upper limit on the 
\rosat\ data taken on 1997 July 23. However, the shape of the decline, shown in
Figure~\ref{lightcurve}, does not depend significantly on $t_{\rm 0}$. This fit 
indicates that the disruption happened approximately three months before the 
\euve\ detection, further strenghtening the association between the UV and
X--ray transient. The slope has a value that is fully compatible with
observations of other candidate TDEs \citep{kom99,cap09,gez09}.

We also fitted the light curve using the \chandra\ data only to understand 1) 
if the assumptions we made to calculate the \euve\ point, obtained by 
extrapolating the UV analysis to the \chandra\ energy range, are correct; 2)
if the \euve\ point is on the same slope of the \chandra\ light curve;
3) if the theoretical models that predict a slope with index $n = 5/3$ at
later time (from days up to few months after the event) are correct. Those 
models predict also that the slope is flatter immediately after the explosion 
\citep{lod09,gez09,str09} with deviations from the $t^{-5/3}$ law more 
pronounced for more centrally concentrated stars (e.g., solar type).

Unfortunately, the large relative errors on the X-ray fluxes do not allow us to 
put firm constrains on $t_{\rm D}$ when fitting only the \chandra\ data. Leaving 
the parameters free to vary, we found that the slope becomes flatter ($n = 1.52 
\pm 0.50$) but $t_{\rm D} = 1999.11 \pm 1.22$, i.e., consistent with a 
pre-\euve\ disruption time within 1$\sigma$, 
although the errors are large. We forced the fit by imposing $t_{\rm D}$ to be 
between the \rosat\ upper limit in 1997 July and the detection by \euve\ in 1998
March. The slope becomes steeper again, ranging from $n = 2.71 \pm 0.36$ when 
$t_{\rm D}$ is fixed immediately after the \rosat\ observation to $n = 2.26 \pm 
0.29$ when $t_{\rm D}$ is set just a few days before the first detection. These 
values are fully compatible with the fit obtained using the \euve\ data. In 
these trials we also calculated the flux at the time of the \euve\ observation 
and found that only by setting $t_{\rm D}$ in the last two months of 1997 will 
the predicted flux (a few $\times 10^{-12}$ \flux) have been compatible with 
what we measured. These results reinforce the idea that 1) the disruption 
event happened a few months before the \euve\ detection; 2) the light curve
peaked before the first observation; 3) the slope of the light curve decay is 
slightly steeper than most simple theoretical predictions.

\section{Discussion}

\subsection{Estimate of the Black Hole Mass}

Based on the photo-redshift, the derived absolute magnitudes indicate that we 
are observing a very dim host, regardless of the exact adopted redshift. If we
make the simplest assumption that the host galaxy lies at the distance of A1795,
we derive an absolute magnitude of $M_{\rm g'} \sim -14.1$ ($M_{\rm B} \sim 
-13.8$, $M_{\rm R} \sim -15.1$). 

Very recently \citet{gra13} have shown that the scaling relation between BH
mass and host spheroid luminosity is bent when core-Sersic and Sersic 
galaxies are considered. The first class contains galaxies whose spheroidal 
component is thought to be created by simple additive dry merger events and is 
more typical for elliptical galaxies, while in the second class the spheroidal 
component is possibly created by gas-rich processes, something more typical of 
spiral galaxies. From the template used to estimate the photo-redshift we 
extrapolated the expected near-IR $K$-band flux (F$_{\rm K} = 14 \pm 5 \mu$Jy 
corresponding to $m_{\rm K} = 19.2 \pm 0.4$). Assuming the cluster distance, the 
absolute magnitude is $M_{\rm K} \sim -18.0 \pm 0.4$. Unfortunately, there are 
no sources included in \citet{gra13} with such a low luminosity. In their 
samples, the core-Sersic galaxies have $M_{\rm K} \simgt -22$ and $M_{\rm B} 
\simgt -18$, while the Sersic galaxies have $M_{\rm K} \simgt -20$ and $M_{\rm B} 
\simgt -16$. As the authors argue, it is not clear if the relation may hold for 
dimmer objects. Furthermore, recent studies \citep{mcg13} show that in 
clusters the central and satellite galaxies may follow distinctly separate 
scaling relations. This can exacerbate the uncertainties in the relation on the
faint end. Since the morphology of this host galaxy is undetermined, we 
tried both 
relations and found, as expected, larger BH mass estimates for the core-Sersic 
cases ($log(M_{\rm BH}/M_{\odot}) = 6.0 \pm 0.9$ and $5.1 \pm 1.0$), with respect
to the Sersic cases ($log(M_{\rm BH}/M_{\odot}) = 2.5 \pm 1.5$ and $2.5 \pm 1.0$).
The two estimates are obtained using the $K$ and $B$ magnitudes, respectively. 
The BH mass in the Sersic cases are very low and this might indicate that the 
scaling relation is not valid for objects with very low absolute magnitude. 
As mentioned earlier, the SED of the host galaxy was best fit using a galaxy 
template with an old, evolved stellar population. This suggests that a 
core-Sersic galaxy is more appropriate. 

In another recent work, \citet{kor13} used more accurate BH masses, partly 
because of improvements in models that include dark matter. They considered only
classical bulges, corresponding to the core-Sersic galaxies defined in 
\citet{gra13}, and ignored the pseudo-bulges, galaxies that fit the Sersic 
definition. Following their proposed $M_{\rm BH} - M_{\rm K,bulge}$ relation we 
found that the mass of the BH is $log(M_{\rm BH}/M_{\odot}) = 5.7 \pm 0.5$. We 
would like to stress that also in this work there are no galaxies below 
$M_{\rm K} \sim -19$ (corresponding to M32, the satellite of the Andromeda 
galaxy) and the relation is not tested at such low luminosities.

The estimated mass is higher if the upper bound in the photo-redshift ($z = 
0.31$) is considered. As mentioned above, the absolute magnitude would decrease
by 3.8 mag, corresponding to $M_{\rm K} \sim -21.8$. For this luminosity, the 
relations for core-Sercic galaxies in \citet{gra13} and classical bulges in 
\citet{kor13} would have given a value for the BH mass of 
$log(M_{\rm BH}/M_{\odot}) = 7.6 \pm 0.5$ and $7.5 \pm 0.5$, respectively.

Combining the results from \citet{gra13} and \citet{kor13} predictions for 
core-Sersic galaxies (classical bulges) we argue that, based on the brightness 
of the host galaxy and the cluster A1795 distance, we are observing an 
intermediate mass BH, with mass in the range $\sim 10^5 - 10^6 M_{\odot}$. 
Throughout the rest of the paper, we assume an average value of $10^{5.5} 
M_{\odot}$.

\subsection{Dynamics}

The analysis of the UV/X-ray light curve did not allow us to put constraints on
the peak time $t_{\rm 0}$. Theoretical models of TDEs predict that after
the encounter half of the stellar debris is unbound from the BH and leaves
the system, while the other half returns to the pericenter after a minimum (or 
fallback) time that is equivalent to the gap between the moment of the 
disruption $t_{\rm D}$ and $t_{\rm 0}$.  This gap depends on the geometry of the 
encounter, the nature of the disrupted star, and the BH mass. Following 
\citet{ulm99} and \citet{mak10}, we define $r_* \equiv R_*/R_{\odot}$ and $m_* 
\equiv M_*/M_{\odot}$, where $R_*$ and $M_*$ are the radius and mass of the 
disrupted star, the penetration factor $\beta = R_{\rm t}/R_{\rm p}$, where 
$R_{\rm p}$ and $R_{\rm t}$ are the periastron and tidal radius of a BH, whose 
mass $M_{\rm 6}$ is in units of $10^6 M_{\odot}$. Then,

\begin{equation}
t_{\rm fallback} = (t_{\rm 0} - t_{\rm D}) = 0.11 k^{-3/2} M_{\rm 6}^{1/2} \beta^3 r_*^{3/2} m_*^{-1} \rm{yr}.
\end{equation}

\noindent While \citet{li02} proposed that the parameter $k$ ranges from 1
for a non-rotating star to a more favorable value of 3 for a star which is spun
up near the point of disruption, more recent work \citep{lod09} suggests that 
the spin-up may not be a significant factor in the fallback evolution. As 
summarized by \citet{mak10}, the factor $r_*^{3/2} m_*^{-1}$ can be simplified as
$m_*^{1/2}$ for main-sequence stars with $M_* < 1 M_{\odot}$ and $m_*^{1/8}$ for 
main-sequence stars with $M_* > 1 M_{\odot}$. Furthermore, the star approaching 
the BH is considered to be on a parabolic orbit, that is $\beta \leq 1$. 
Assuming the most conservative value for the penetration factor $\beta = 1$, a 
mass for the main sequence star in the range 0.1--100 $M_{\odot}$, a spin value 
in the range of $(1 ; 3)$, and $log(M_{\rm BH}/M_{\odot}) = 5.5 \pm 0.5$, 
then $(t_{\rm 0} - t_{\rm D}) \in (0.002 ; 0.19)$ yr, corresponding to 
$(0.8 ; 70)$ days.

Our light curve analysis shows a steeper decay than typically expected for a 
TDE case (the canonical $n = 5/3$ power-law). Initially, \citet{lod11}
proposed a more rapid decay for the X-ray light curve at late times due to a 
change in the spectral behavior caused by a drop in the blackbody temperature.
Our X-ray spectral analysis suggests no such spectral evolution. A 
different possibility is that the light curve shape is affected by the dynamic 
of the encounter \citep{can11,gui13}. According to the hydrodynamical 
simulations of \citet{gui13}, the peak will occur at

\begin{equation}
t_{\rm peak} = (t_{\rm 0} - t_{\rm D}) = B_{\gamma} M_{\rm 6}^{1/2} r_*^{3/2} m_*^{-1} \rm{yr}, 
\end{equation}

\noindent while the asymptotic light curve slope decay is

\begin{equation}
n_{\infty} = D_{\gamma}, 
\end{equation}

\noindent where $B_{\gamma}$ and $D_{\gamma}$ are functions of the penetration 
factor $\beta$, with different behavior if the polytropic index $\gamma$ is 
$4/3$ or $5/3$, values assumed for high and low mass main sequence stars, 
respectively. From the 
light curve analysis $n = n_{\infty} = 2.44\pm0.40$, this means that $\beta$ must
be in the range 0.6--0.8 if $\gamma = 5/3$ and 0.7--1.7 if $\gamma = 4/3$. 
Consequently, $(t_{\rm 0} - t_{\rm D}) \in (19 ; 72)$ and $(9 ; 66)$ days for the 
two stellar structures, respectively. The latest interval can be further 
shortened by considering the fastest decay in \citet{gui13}: their steeper decay
$n_{\infty} \sim 2.2$ is obtained with $\gamma = 4/3$ and $\beta \in (0.9 ; 
1.6)$, corresponding to $(t_{\rm 0} - t_{\rm D}) \in (16 ; 25)$ days. A much 
shorter time-scale is obtained assuming the canonical $n = 5/3$ decay, an 
average BH mass of $10^{5.5} M_{\odot}$ and a solar-type star ($m_* = r_* = 1$):
we find $(t_{\rm 0} - t_{\rm D}) = 0.012$ yr (i.e., 4 days).

Since $t_{\rm D} = 1997.957$, we find that $t_{\rm 0}$ very likely happened before
the observation by \euve\ (1998.236) both using the canonical or the new 
theoretical assumptions, confirming the results from the light curve analysis.

\subsection{Energetics}

The analysis of the X-ray spectrum indicates that only thermal emission is 
necessary. The most obvious interpretation is that we are seeing emission from 
an accretion disk generated by a TDE. Assuming a thermal, black body model with
$kT = 0.09$ keV as the best description of the X-ray spectrum over the duration
of the flare, the factor to convert the 0.5--7 keV into bolometric luminosity is
10.5. The conversion factor might be slightly higher if the X-ray spectrum is 
more complex, as possibly seen in the first \chandra\ observation in 1999. 
Despite the 
good fit obtained with a power-law in that observation, we do not think that 
this is the correct description of the photon spectral energy distribution: by 
extrapolating the unabsorbed flux from \chandra\ into the \euve\ energy range at
the time of the \euve\ observation in 1998 using the power-law model we found an
absorbed, observed flux of $1.6 \times 10^{-11}$ \flux, which is $\sim4 
\times$ what we measured.

We estimate the total released energy by integrating the light curve over the
course of the flare. \citet{lod11} showed that the bolometric luminosity light 
curve for a $10^6 M_{\odot}$ BH and solar-type star is compatible with the 
typical decay ($n = 5/3$) only $\sim 200$ days after the event, while it departs
from that decay at earlier times: at the time of the peak \citet{lod11} predict 
a bolometric luminosity $\sim 4\times$ lower than if the 5/3 decay is assumed. 
We started integrating the light curve at the earliest point $(t_{\rm 0} - 
t_{\rm D}) = 0.012$ yr (see above) and we stopped at the last detection by 
\chandra\ (on 2004 January 18). We found a value for the total released 
energy of $E = 1.7 \times 10^{52}$ erg at the cluster redshift. Assuming a 
standard mass-to-energy conversion factor $\epsilon=0.1$, the mass accreted over
the 6 year time frame is then $M_{\rm acc} = E/(\epsilon c^2) \sim 0.10 
M_{\odot}$. \citet{aya00} showed through numerical simulations that for a black 
hole with $10^6 M_{\odot}$ mass and a solar-type disrupted star, only $\sim$10\% 
of the star is accreted, i.e., the assumption of a solar-type star in our case 
is consistent with these results.

The bound mass forms a disk and accretes on the BH initially at a high rate. 
\citet{ree88} and \citet{phi89} showed that the mass accretion rate is 

\begin{equation}
\dot{M}_{\rm fallback} \approx \frac{1}{3} \frac{m_*}{t_{\rm fallback}} \left(\frac{t}{t_{\rm fallback}}\right)^{-5/3}.
\end{equation}

\noindent At the fallback time (4 days), the accretion rate was $\sim 28 
M_{\odot}$ yr$^{-1}$ but it fell quickly and at the time of the \euve\ 
observation ($\sim 100$ days), the rate was $\sim 0.1 M_{\odot}$ yr$^{-1}$. For a
$10^{5.5} M_{\odot}$ BH mass, the Eddington accretion rate is $\dot{M}_{\rm Edd} 
\equiv 10L_{\rm Edd}/c^2 = 7.0 \times 10^{-3} M_{\odot}$ yr$^{-1}$, where 
$L_{\rm Edd}$ is the Eddington luminosity ($4.1 \times 10^{43}$ \lum), and 
0.1 is the efficiency.
This means that up to the \euve\ observation, the accretion rate was still 
super-Eddington. During this phase the formed disk is thought to be 
geometrically thick, optically thin and highly advective \citep{kin03}. The 
change from super- to sub-Eddington rate happens at $t_{\rm Edd} \sim 0.1 
M_{\rm 6}^{2/5} R_{\rm p,3R_S}^{6/5} m_*^{3/5} r_*^{-3/5}$ yr \citep{str09}, where 
$R_{\rm p,3R_S}$ is the pericenter distance in units of 3 Schwarzschild radii, 
$R_{\rm S}$. Assuming $M_{\rm BH} =  10^{5.5} M_{\odot}$, a solar-type star, and 
$t_{\rm Edd} \sim 1.7$ yr (from Eq. 4), this implies that $R_{\rm p,3R_S} \sim 
14.7$, or $\sim 4.1 \times 10^{12}$ cm. Since $R_{\rm t} = r_* m_*^{-1/3} 
M_{\rm BH}^{1/3} = 4.8 \times 10^{12}$ cm, the penetration factor is $\beta \sim 
1.2$.

The results are only slightly different if the model of \citet{gui13} and the
assumptions explained above ($M_{\rm BH} =  10^{5.5} M_{\odot}$, $\gamma = 4/3$ 
and $\beta \in (0.9 ; 1.6)$) are used: The accretion rate at the time of the 
peak (between 16 and 25 days) varies from 0.36 to 4.8 $M_{\odot}$ yr$^{-1}$, 
while it becomes sub-Eddington at sometimes between 2.9 and 3.5 yr after the 
disruption. 

\citet{li02} showed that the accretion disk has a characteristic radius 
($R_{\rm X}$) that might be estimated from the X-ray spectral analysis. By 
requiring a black body model with temperature $T_{\rm bb}$ and assuming a 
correction factor of $f_{\rm c} \geq 1$ \citep{ros92} to compensate for spectral 
hardening by Comptonization and electron scattering, the radius can be expressed
as 

\begin{equation}
R_{\rm X} = \left( \frac{L_{\rm bol} f_{\rm c}^4}{\pi \sigma T_{\rm bb}^4} \right)^{1/2},
\end{equation}

\noindent where $L_{\rm bol}$ is the bolometric luminosity and $\sigma$ is 
the Stefan-Boltzmann constant. Since our \chandra\ spectra were fit with a 
blackbody model with $T_{\rm bb} = 10^6$ K, and at the time of the \euve\ 
observation $L_{\rm bol}$ was $3.3 \times 10^{44}$ \lum, we find that $R_{\rm X}$ 
is $ \geq 1.36 \times 10^{12}$ cm. This radius coincides with the tidal radius 
if a correction factor $f_{\rm c} \sim 2$ is assumed, a value similar to what 
has been found
appropriate for these conditions \citep[i.e., $f_{\rm c} = 3$, ][]{shi93,li02}.
$R_{\rm X}$ is where the inner edge of the debris stream and the corresponding 
elliptical disk should be located, assuming that the debris stream is centered 
at $R \sim 2 \times R_{\rm p}$. The radius does not correspond to the innermost
stable circular orbit ($R_{\rm ISCO}$) that is expected to mark the inner edge of
the accretion disk. Since $R_{\rm ISCO} = 3 \times R_{\rm S}$ for a 
non-spinning black hole, it is at least an order of magnitude smaller than the 
estimated $R_{\rm X}$.

\section{Summary}

We serendipitously discovered a high energy transient in the field of view of 
the moderately rich cluster Abell 1795. The flare was discovered in observations
with the \euve\ and \chandra\ satellites: the first detection was on 1998 March 
27 in the UV and the last glimpse was on 2004 January 18 in the X-rays. Previous
observations by \euve\ and \rosat\ up to 1997 July do not reveal any emission at
the transient position. A total of 7 observations are used to generate the X-ray
light curve: the brightness of CXO J1348 decays as $a \times 
[(t-t_{\rm D})/(t_{\rm 0}-t_{\rm D})]^{-n}$, where $n = 2.44\pm0.40$, a behavior 
seen in previous cases of TDE candidates \citep{cap09} and in agreement with 
recent hydrodynamical simulations \citep{gui13}. The start of this event 
$t_{\rm D}$ can be set a few weeks before the first observation. The spectral 
analysis of the \chandra\ data are consistent with this interpretation. There is
not significant emission in the hard X-rays (above 2-3 keV) associated with this
event, and the fit of the soft spectrum can be obtained
using a thermal model. The lack of any flat power-law component rules out some 
of the candidate progenitors, such as AGN and GRBs. Also a supernova explosion 
is excluded because of the very short-term variability observed by \chandra\ 
over a 4 day time range, as well as the lack of star formation in the host. 
Throughout the first 3 years of \chandra\ observations
the temperature of the blackbody model is constant around $kT=0.09$ keV, 
corresponding to $T = 10^6$ K. Assuming this model to fit the \euve\ observation
as well, we estimated that the 0.5--7 keV unabsorbed flux changes by a factor of
$\sim2300$, a value higher than numerous other TDEs and consistent with the
most extreme cases monitored over very long periods \citep[e.g., \chandra\ 
observations of RX J1624.9+7554 showed a decline by a factor of 6000 from its 
\rosat\ peak; ][]{hal04}.

At the position of CXO J1348 we found a host galaxy using ground- and 
space-based telescopes. Combining archival with new and deeper observations, we 
were able to estimate a photo-redshift. The best fit of the SED can be obtained 
with an elliptical/S0 galaxy located in proximity of A1795. The estimated 
absolute magnitude place this galaxy at the very bottom of previously studies of
galaxies containing either a SMBH or a compact nucleus. Using different methods 
\citep{gra13,kor13} we concluded that the BH in the center of the host might not
be super-massive ($10^{5.5\pm0.5} M_{\odot}$). Only a handful of cases has been 
reported so far in which an IMBH has been found in the host galaxy (see 
Sect.~1). \citet{wan04} predicted that disruptions happen with a rate of $\sim
10^{-4}-10^{-5}$ galaxy$^{-1}$ yr$^{-1}$ for galaxies with supermassive BH, while
the rate is highest in nucleated dwarf galaxies, reaching a value as high as 
$\sim 10^{-3}$ galaxy$^{-1}$ yr$^{-1}$ in faint nucleated spheroids. 
During the course of 12 years of observations of A1795, \chandra\ covered 
exensively the cluster inner part (10\arcmin\ in radius). In that area there 
are more than 2000 objects that have an optical magnitude between 21 and 24, a 
plausible range for dwarf spheroids at the cluster distance. Assuming that only
10\% of those objects are located at that distance, than the measured rate is 
also in the range $\sim 10^{-3}-10^{-4}$. Thus, our serendipitous discovery can
be explained by this high disruption rate and argues in favor the existence of 
an IMBH in a dwarf galaxy.

The host shows no significant variability over a very wide time range (from 
1999 to 2013), while an optical spectrum obtained with the LBT 15 years after 
the event does not show the presence of emission lines, supporting the idea that
the optical emission is dominated by the underlying quiescent galaxy. Given the 
high temperature observed in the X-ray spectrum, the optical transient wouldn't 
be observed, even at the time of the X-ray peak: the estimated flux at 5500 \AA\
assuming a simple blackbody model would have been $\sim0.02$ $\mu$Jy, 
corresponding to mag$_V$=28. It has been shown \citep{str09,lod11} that the
optical radiation might be dominated by emission generated in wind outflows 
during the super-Eddington phase. This enhancement can be up to two orders of
magnitudes, corresponding to mag$_V$=23, similar to the value of the entire
host galaxy. Only with observations performed within a few months after the 
event it would have been possible to detect this increase. Unfortunately, the 
first available observation, obtained by \hst, happened more than a year after 
the disruption and no enhancement has been observed.

Almost all the known TDE cases are not detected in the radio bands and have
a X--ray luminosity of $L_{\rm X} \simlt 10^{45}$ \lum\ \citep{kom02}. The only 
exceptions are Sw J1644+57 and Sw J2058+05, whose detection at low frequencies 
and large luminosity in the X-rays 
\citep[$L_{\rm X} > 10^{47}$ \lum,][]{can11,bur11} suggest the presence of a jet 
of relativistic, collimated plasma which produces synchrotron emission and
boosts the luminosity by 
beaming. In our case the extrapolated bolometric luminosity is $2.8 \times
10^{48}$ \lum\ at the time of the theoretical peak, 2.4 days after the event. 
\citet{lod11} showed that this is likely a very conservative upper limit
because the light curve is flatter in the first part of the system evolution.
The lack of any radio signal in any observations from 2000 to 2005 indicates 
that no jet has been created or that the direction of the collimated plasma does
not coincide with the line of sight. It has been proposed and shown 
\citep[e.g.,][]{gia11,van11,bow13} that radio emission can
be visible one or more years after the TDE, when the plasma decelerates to 
mildly relativistic speed due to interaction with the interstellar medium or 
when the jet becomes radio-loud as a function of the accretion rate. New radio 
observations are needed to check if this assumption applies to this case.

This study highlights, once again, the importance of an X-ray monitoring 
campaign of clusters of galaxies to discover and better characterize this kind 
of event. As shown above, the lack of information in the very early part of the
event prevented us to put firm constraints to the system. Future monitoring 
programs in the X-rays, such as {\it eROSITA} \citep[e.g.,][]{pre07}, would
be extremely important to find other TDEs associated with IMBHs, whose 
existence is still open to debate.

\acknowledgments
We thank the anonymous referee for his/her constructive comments, which helped us to improve the manuscript.

Observations have been carried out using the Large Binocular Telescope (LBT) at
Mt. Graham, AZ., the Johnson Telescope at the Observatorio Astrono\'mico 
Nacional/San Pedro M\'artir in Mexico, and the Nordic Optical Telescope at La 
Palma, Spain.

The LBT is an international collaboration among institutions in the United 
States, Italy, and Germany. LBT Corporation partners are The University of 
Arizona on behalf of the Arizona university system; Istituto Nazionale di 
Astrofisica, Italy; LBT Beteiligungsgesellschaft, Germany, representing the 
Max-Planck Society, the Astrophysical Institute Potsdam, and Heidelberg 
University; The Ohio State University; and The Research Corporation, on behalf 
of The University of Notre Dame, University of Minnesota, and University of 
Virginia. This paper used data obtained with the MODS spectrographs built with
funding from NSF grant AST-9987045 and the NSF Telescope System Instrumentation 
Program (TSIP), with additional funds from the Ohio Board of Regents and the 
Ohio State University Office of Research.

We thank the staff of the Observatorio Astron\'omico Nacional on Sierra 
San Pedro M\'artir and the RATIR instrument team 
(ratir.astroscu.unam.mx). RATIR observations with the Harold L. Johnson 
1.5-meter telescope of the  Obervatorio Astron\'omico Nacional on Sierra 
San Pedro M\'artir are partially funded by UNAM, CONACyT and NASA, and 
supported by the loan of an H2RG detector by Teledyne Scientific and 
Imaging.

The data presented here were obtained in part with ALFOSC, which is provided by
the Instituto de Astrofisica de Andalucia (IAA) under a joint agreement with the
University of Copenhagen and NOTSA.

C.C.C.\ was supported at NRL by a Karles' Fellowship and NASA DPR S-15633-Y.

H.L. acknowledges financial support by the European Union through the COFUND 
scheme.

We thank Marco Fumana from the LBT data center for quickly providing the 
reduced observations; Martin Sirk for helping in the \euve\ analysis; Dan Perley
for estimating the photo-redshift; and 
Sjoert van Velzen for providing additional comments on the discussion.

{\it Facilities:} \facility{EUVE (DS)}, \facility{HST (WFPC2)}, 
\facility{CXO (ACIS)}, \facility{ROSAT (PSPC)}, \facility{VLA}, \facility{VLT 
(FORS1)}, \facility{LBT (MODS)}, \facility{OAN/PSM (RATIR)}, \facility{NOT}.

\clearpage

\clearpage

\clearpage

\clearpage

\begin{table}
\begin{center}
\caption{Observation Log of \chandra\ Pointings through 2005. \label{xray}}
\begin{tabular}{cccccc}
\tableline\tableline
ObsID & Date        & ACIS & Exp. & c/r\\
 (1)  &     (2)     & (3)  & (4)  & (5) \\
\tableline
00494 & 1999 Dec 20  & S  &  19.5  &  10.5$\pm$0.9 \\
00493 & 2000 Mar 21  & S  &  19.6  &   7.6$\pm$0.7 \\
03666 & 2002 Jun 10  & S  &  14.4  &   1.9$\pm$0.4 \\
05287 & 2004 Jan 14  & S  &  14.3  &  $<$0.5       \\
05288 & 2004 Jan 16  & S  &  14.6  &   1.4$\pm$0.4 \\
05289 & 2004 Jan 18  & I  &  15.0  &   1.3$\pm$0.3 \\
06160 & 2005 Mar 20  & S  &  14.8  &  $<$0.5       \\
06162 & 2005 Mar 28  & I  &  13.6  &  $<$0.6       \\
06163 & 2005 Mar 31  & I  &  14.9  &  $<$0.6       \\
\tableline
\end{tabular}
\tablecomments{
{\bf Column explanations}: 1=Observation ID; 2=Observation Date; 3=Instrument 
where the position of CXO J1348 is localized; 4=Exposure time in ksec; 
5=Net count rates for detections in units of $10^{-3}$ ct s$^{-1}$ in the 0.3--8 
keV range.
}
\end{center}
\end{table}

\clearpage

\begin{table}
\begin{center}
\caption{Observation Log and Point Source $3\sigma$ Limits from the VLA. \label{radio}}
\begin{tabular}{lccccc}
\tableline\tableline
Date & Program & Freq. & Beam & Exp. & Flux Density  \\
(1)  & (2)     & (3)   & (4)  & (5)  & (6)    \\
\tableline
2000 Oct 05 &    AP405	& 8.46  &    11.50, 9.05,  $-$66.3 & 590 &    $<$0.10 \\
2003 Jul 11 &    AF403	& 4.71  &    0.391, 0.446, $-$33.5 & 320 &    $<$0.32 \\
2005 Oct 24 &    AL663	& 8.46  &    15.70, 5.55,  $-$72.1 & 63  &    $<$0.13 \\
2005 Oct 24 &    AL663	& 4.86  &    9.62,  3.05,  $-$72.5 & 83  &    $<$0.18 \\
\tableline
\end{tabular}
\tablecomments{
{\bf Column explanations}: 1=Observation date; 2=Program; 3=Frequency in GHz; 4=Gaussian restoring beam dimensions are the major axis (\arcsec), minor axis (\arcsec), and the position angle in degrees; 5=Exposure time in seconds; 6=Detection limit ($3\sigma$) in mJy.
}
\end{center}
\end{table}

\clearpage

\begin{table}
\begin{center}
\caption{Observation Log and Photometry in the UV/Optical/near-IR Bands. \label{optical}}
\begin{tabular}{llllrcl}
\tableline\tableline
Telescope  & Date  & Instrument & Filter & Exp. & Magnitude & System\\ 
 (1)       &  (2)  & (3)        &  (4)   & (5)  & (6)    & (7) \\ 
\tableline
HST     & 1999 Apr 11  & WFPC2     & F555W      &  300  & 23.02$\pm$0.13 & V  \\ 
VLT & 2002 Jun 29\tablenotemark{a} & FORS1 & U  &  2840 & 23.43$\pm$0.28 & V  \\ 
        &              &           & B          &  1480 & 23.61$\pm$0.10 & V  \\ 
        &              &           & R          &  800  & 22.14$\pm$0.05 & V  \\ 
CFHT    & 2008 Aug 05  & MEGAPRIME & $g'$       &  240  & 23.11$\pm$0.05 & AB \\ 
        &              &           & $r'$       &  120  & 22.22$\pm$0.05 & AB \\ 
        & 2009 Jul 25  & MEGAPRIME & $g'$       &  240  & 23.00$\pm$0.05 & AB \\ 
        &              &           & $r'$       &  120  & 22.15$\pm$0.05 & AB \\ 
INT     & 2010 May 10  & WFC       & U          &  400  & 23.59$\pm$0.16  & V  \\ 
        &              &           & B          &  400  & 23.42$\pm$0.09  & V  \\ 
        &              &           & V          &  400  & 22.37$\pm$0.06  & V  \\ 
NOT     & 2012 Mar 20  & ALFOSC    & B          &  1500 & 23.40$\pm$0.15 & V  \\ 
        &              &           & $i'$       &  900  & 21.40$\pm$0.15 & AB \\ 
        & 2013 Mar 14  & MOSCA     & U          &  6000 & 24.60$\pm$0.30 & AB \\ 
Johnson & 2013 Feb 12  & RATIR     & $H$        & 1320  & 19.68$\pm$0.35 & V  \\ 
        & 2013 Feb 19  &           & $g$        & 4800  & 23.14$\pm$0.20 & AB \\ 
        &              &           & $r$        & 4800  & 22.16$\pm$0.08 & AB \\ 
        &              &           & $i$        & 9600  & 21.91$\pm$0.06 & AB \\ 
        &              &           & $Z$        & 7200  & 21.58$\pm$0.12 & AB \\ 
        &              &           & $J$        & 7200  & 20.28$\pm$0.12 & V  \\ 
\tableline
\end{tabular}
\tablenotetext{a}{This a sum of observations with equal exposures taken on 2002 June 08 and July 19}
\tablecomments{ 
{\bf Column explanations}: 1=Telescope; 2=Observation date; 3=Instrument or 
camera; 4=Filter; 5=Exposure time in seconds; 
6=Magnitude; 7=Photometric system (Vega or AB).
The observations are ordered by date and frequency, starting with the bluer filter.}
\end{center}
\end{table}

\end{document}